\documentstyle[]{article}
\font\cero=cmss10 scaled 1728 \font\uno=cmssbx10 scaled 1200
\begin{document}
\begin{flushleft}
{ \cero Hamiltonian dynamics for  Einstein's action in   G$\rightarrow$0 limit  }  \\ [3em]
\end{flushleft}
{\sf Alberto Escalante}  \\
 {\it 
 Instituto de F{\'i}sica Luis Rivera Terrazas, Benem\'erita Universidad Aut\'onoma de Puebla, (IFUAP). \\
   Apartado postal      J-48 72570 Puebla. Pue., M\'exico\\
   LUTh, Observatoire de Paris, Meudon, France, \\
 } 
  (aescalan@sirio.ifuap.buap.mx, alberto.escalante@obspm.fr) \\[4em]

\noindent{\uno Abstract} \vspace{.5cm}\\
The Hamiltonian analysis for the Einstein's action in $ G\rightarrow0 $  limit is performed. Considering the original configuration space without involve the usual $ADM$ variables we show that the version $ G\rightarrow0 $ for  Einstein's action is devoid of physical degrees of freedom.  In addition,  we will  identify the relevant symmetries of the theory such as the extended action, the extended Hamiltonian, the gauge transformations and the algebra of the constraints. As complement part of this work, we develop the covariant canonical formalism where will be constructed a closed and gauge invariant symplectic form. In particular, using the geometric form we will obtain by means of other way the same symmetries that we found using the Hamiltonian analysis.

\begin{center}
{\uno I. INTRODUCTION}
\end{center}
\vspace{1em} \
Hamiltonian analysis for   Einstein's theory of gravity has  been great topic of study   in the last  years. As we know,  the history begins with the work  reported  by Arnowitt-Deser-Misner ($ADM$)  where  the  $3+1$ split  of the space time   allows us to study  the Hamiltonian dynamics, the constraints and the symmetries of general relativity theory. In the $ADM$ work, the fundamental variables to preform the Hamiltonian analysis  are considered the 3-metric  and its  respectively conjugate momenta  \cite{1}. However,  when we try to make progress in the quantization of   the theory  this program  presents  difficulties,    because the no linearly of the gravitational field is manifested  in the  constraints. In this manner,  at quantum level  to work with these variables (ADM variables)  presents  several problems.  \\
In the 80's,  the panorama becomes to be clarified thanks to the greats  works developed  by Ashtekar introducing a kind of new variables  for studying  the Hamiltonian dynamics for the gravitational field \cite{2,3,4}. The use of these new variables leads to a important  simplification of the equations of the theory. In this program, both the constraints and the evolution equations of the canonical general relativity become simple polinomials of the field variables. Nevertheless, the price to pay for these simplifications   is that the Astekar's variables are complex, and therefore Ashtekar canonical formulation describes complex general relativity.  In order to obtain the real physical degrees of freedom  one needs  to append a posteriori appropiate reality conditions \cite{5,6}. After  the Asthekar's works,  the study of canonical gravity in its classical or quantum form  has been of great  interest in the literature \cite{7, 8, 9, 10, 11, 12, 13},  especially in  the loop quantum gravity  context \cite{14, 15}.\\
On the other hand,  in  recently works  has been proposed  to study using the Ashtekar formulation   the    $ G\rightarrow0 $ limit of Euclidean or complexified  general  relativity,   where  the quantization of the theory in the loop representation is obtained  and infinite dimensional space of exact solutions to the constraints are found \cite{16}.  The study of Einstein's theory in this limit becomes to be  relevant because we could make progress to study a different approach to perturbation theory at quantum level. As we know,  the standard way for  studying  this important part in gravity is making  the perturbation  around a classical background metric,   but  in the process the    relevant symmetries of Einstein's theory are lost, namely  the background independence and diffeomorphisms.  However,  the model reported in  \cite{16}     marks    a big difference respect to the standard treatment because in the limit   the  symmetries of general relativity are not lost. Thus,   we could  have  now  a  new starting point  to analyze  in the mentioned limit  a full diffeomorphism invariant  and background independent theory. \\  
On the other side, in this same context we find in \cite{17}  other  different proposal,   where  setting  the  $ G\rightarrow0 $  limit  for general relativity  written in the first order formalism  and under a  change of variables, the theory becomes to be  a  copy of  abelian $BF$ topological field theory. Furthermore, using a kind of ($ADM$) variables  the Hamiltonian analysis for  the theory  is  performed, allows us to find a connection with parametrized field theory \cite{17, 18}. It is important to observe  that the models purposed in \cite{16} and \cite{17} are quite different. In the first one model,  the    Astekar's variables has been used  and the  relevant results  reported  are that Euclidean general relativity in the $ G\rightarrow0 $ limit  is not a  free theory because  the model has two degrees of freedom. In the second one model,  we  find that in $ G\rightarrow0 $ limit   general relativity expressed  in the first oder formalism  becomes to be a free field theory.   \\
With  all these antecedents, the purpose of this paper is to report the Hamiltonian analysis  for the model presented in \cite{17} without involve the $ADM$ variables. The reason to do  this  is simple, we wish  to report  the symmetries and the constraints of the theory from other point of view. This is, in this work  we report  the Dirac's analysis using only the dynamical variables  implicated  in the action. In this way,  we are showing that is possible to  obtain the same physical  information for the theory  without resort to  ADM variables.  We  finish our analysis  developing   the covariant canonical formalism  for the theory under study, and we obtain by means of a different way the symmetries found using the Hamiltonian method. Therefore,  in this work   we are establishing  the bases to quantize the theory in forthcoming works.\\
The paper is organized as follows. In  Section II,  we present a pure Dirac analysis for general relativity in $ G\rightarrow0 $ limit. As important part that we will  find   in this section  are the extended action,  the extended Hamiltonian and  the identification of the  first and second class constraints. In addition, with the complete classification of the constraints we carry out the counting of the physical degrees of freedom and we   present the Dirac bracket for the theory. In Section II.I,  using  Catellani's algorithm we will find   the gauge symmetries for the theory. In particular we we prove  that the theory under study is invariant under diffeomorphisms. In Section III,  using basic concepts of symplectic geometry we construct a closed and gauge invariant  symplectic form on the covariant phase space,  which turns represent a complete covariant canonical  description of the theory. Using the present geometric form, we reproduce the results found with the Hamiltonian method.   In Section IV, we give some conclusions and prospects . \\
\newline
\newline
\setcounter{equation}{0} \label{c2}
\noindent \textbf{II. Hamiltonian analysis}\\[1ex]
As we know,  the Einstein's action for gravity written in the first order formalism is expressed  by \cite{14, 16}
\begin{equation}
S[e, \omega]=\frac{1}{4}\int_M \epsilon^{IJKL} e_I \wedge e_J \wedge R_{KL} [\omega ]  ,
\label{eq1}
\end{equation}
where $e^I= e^I{_{\mu}} dx^{\mu}$ is the one-form tetrad field,   $R^{IJ}[\omega]= \frac{1}{2}  R^{IJ}{_{\mu \nu}} dx^\mu \wedge dx^\nu$
is the curvature  of the $SO(3,1)$ 1-form  connection $\omega_{\nu}{^ {IJ}}$ with  $R^{IJ}{_{\mu \nu}}= \partial_\mu \omega_{\nu}{^ {IJ}} - \partial_\nu \omega_{\mu}{^ {IJ}} + G (\omega_{\mu}{^ {IK}}\omega_{\nu}{_K{^J}}-\omega_{\nu}{^ {IK}}\omega_{\mu}{_K{^J}} )$. Here, $G$ is the gravitational coupling constant, $\epsilon^{IJKL}$ is the completely antisymmetric object with $\epsilon^{0123}=1$,   $\mu, \nu=0,1,..,3$ are spacetime indices, $x^\mu$  are the coordinates that label the points fo the 4-dimensional manifold $M$ and  $I, J= 0,1..,3$ are internal indices that can be raised and lowered by the internal Lorentzian   metric  $\eta_{IJ}= (-1,1,1,1)$.\\
Setting the  $G\rightarrow 0$ limit ,  the above action becomes to be 
\begin{equation} 
S[e,\omega]= \frac{1}{8}\int_M \epsilon^{\alpha \beta \mu \nu } \epsilon^{IJKL} e_{I \alpha} e_{J \beta} (\partial_\mu \omega_{\nu}{^ {IJ}} - \partial _\nu \omega_{\mu}{^ {IJ}}) dx^4.  
\label{eq2}
\end{equation}
where $\epsilon^{\alpha \beta \mu \nu }$  is the volume 4-form. Calculating  the variation of the action (\ref{eq2})  we find the next equations of motion 
\begin{equation}
\epsilon^{\alpha \beta \mu \nu }\partial_{[\mu} e_{\nu]I} =0,
\label{eq3}
\end{equation}
and
\begin{equation}
 \epsilon^{\alpha \beta \mu \nu } \partial_\mu B_{I \alpha \beta}=0,
 \label{eq4}
\end{equation}
 here, the two-forms $B{^{I}}_{\alpha \beta}$ are defined by  $B{^{I}}_{\alpha \beta}= - \frac{1}{2} \epsilon^{IJKL} e_{[\alpha J} \omega_{\beta] KL}$,  provided  that the tetrad is non-degenerate,  $B^I$ has  inverse $\omega_{\alpha IJ}= \frac{1}{2} \epsilon_{IJKL} e^{\beta K} \left( B{^{L}}_{\alpha \beta} - \frac{1}{2} e^{\gamma L} e_{\alpha N} B{^{N}}_{\beta \gamma}   \right)$. We can see that equation (\ref{eq3}) implies that   $e_{\alpha I}= \partial_\alpha f_I$, so  $g_{\mu \nu}=\eta_{IJ}\partial_\mu f^I\partial_\nu f^J$. Which corresponds to (locally) Minkowski spacetime \cite{17}.  \\
With  all these preliminar results,  using the variable $B$ and integrating by parts we can rewrite the  action (\ref{eq2})  in the next form 
\begin{equation}
S[B, e]= \frac{1}{2}\int_M \epsilon^{\alpha \beta \mu \nu} B{^{I}}_{\alpha \beta } (\partial_\mu e_{\nu I}- \partial_{\nu}e_{\mu I} )dx^4.
\label{eq5}
\end{equation}
Thus, we can obtain from (\ref{eq5})  the same equations of motion given in (\ref{eq3}) and (\ref{eq4})  considering to $B$ and $e$ as our new dynamical variables. It is remarkable to note that  the action (\ref{eq1}) which has an $SO(3,1)$ connection $\omega_{\nu}{^ {IJ}}$,   in the  $G\rightarrow0$ limit    (\ref{eq2}) becomes to be a collection of six $U(1)$ connections and the tetrad field $e^I{_\mu}$ is a collection of four gauge invariant vector fields, we will prove this point  performing the Hamiltonian analysis in the next lines.\\ 
The starting point of this work is the action (\ref{eq5}),  but  to difference of the paper reported in \cite{17} we will not involve  a kind of $ADM$ variables for  performing  the Hamiltonian analysis,  in spite of   in the canonical gravity  context   the  standard way  for  developing the Hamiltonian dynamics is using these  variables. The reason to do this  is because in this work we aim to report  the Dirac's method working with the full configuration space, this is, we will develop the Dirac analysis using only the   configuration variables involved  in the action (\ref{eq5}), namely  $B,e$. In this way, we can  know the  constrains in his complete form without fix any  gauge, the symmetries, the extended action and the extended Hamiltonian for the theory. Of course, if we wish we can obtain the results reported by Nuno {\it et. al} \cite{17}  as particular case of this paper considering  the second class constraints as strong equations.  Thus, with this letter we  are establishing  the basis to quantize  the theory  described by (\ref{eq5}) which will be reported in forthcoming  works.     \\
By performing the 3+1 decomposition in the action (\ref{eq5}) we find
  \begin{equation}
S[B,e]= \int  \left[  \eta^{abc} B_{I ab} \dot{e}{^{I}}_c + \frac{1}{2} \eta^{abc} B_{I0a} (\partial_b e{^{I}}_{c}-\partial_c e{^{I}}_b ) - (\eta^{abc}B_{Iab } )  \partial_ce{^{I}}_0 \right] dx^4, 
\label{eq6}
\end{equation}
where $ \eta^{abc}=\epsilon^{0abc}$, $a,b,c=1,2,3$. From (\ref{eq6}),  we can identify the Lagrangian density given by 
\begin{equation}
{\mathcal{L}}=  \eta^{abc} B_{I ab} \dot{e}{^{I}}_c + \frac{1}{2} \eta^{abc} B_{I0a} (\partial_b e{^{I}}_{c}-\partial_c e{^{I}}_b )  -(\eta^{abc}B_{Iab } )  \partial_ce{^{I}}_0. 
\label{eq7}
\end{equation}
Dirac's method calls for the definition of the momenta $(\Pi{_{I}}^{\alpha \beta}, \Pi{_{I}}^{\alpha}) $ canonically conjugate to  $(B{^{I}}_{\alpha \beta}, e{^{I}}_{\mu})$ \cite{19} 
\begin{equation}
\Pi{_{I}}^{\alpha \beta}= \frac{\delta {\mathcal{L}} }{ \delta \dot{B}{^{I}}_{\alpha \beta} }, \quad \quad  \Pi{_{I}}^{\alpha}=  \frac{\delta {\mathcal{L}} }{ \delta \dot{e}{^{I}}_{\mu} }, 
\label{eq8}
\end{equation}
on the other hand,  the matrix elements of the  Hessian 
\begin{equation}
\frac{\partial^2{\mathcal{L}} }{\partial (\partial_\mu B{^{I}}_{\alpha \beta } ) \partial(\partial_\mu B{^{J}}_{\rho \sigma } ) }, \quad \frac{\partial^2{\mathcal{L}} }{\partial (\partial_\mu e{^{I}}_{\alpha } ) \partial(\partial_\mu B{^{J}}_{\rho \sigma } ) }, \quad \frac{\partial^2{\mathcal{L}} }{\partial (\partial_\mu e{^{I}}_{\alpha } ) \partial(\partial_\mu e{^{J}}_{\beta } ) },
\label{eq9}
\end{equation} 
are identically zero,  the rank of the Hessian is zero. Thus,  we expect 40 primary constraints. From the definition of the momenta  (\ref{eq8}) we identify  the next 40 primary constraints 
\begin{eqnarray}
\phi{_{I}}^{0}&:=& \Pi{_{I}}^{0} \approx 0 ,\nonumber \\
\phi{_{I}}^{a}&:=& \Pi{_{I}}^{a} - \eta^{abc}B_{I bc} \approx 0, \nonumber \\
\phi{_{I}}^{0a}&:=& \Pi{_{I}}^{0a} \approx 0,  \nonumber \\
\phi{_{I}}^{ab}&:=& \Pi{_{I}}^{ab} \approx 0.
\label{eq10}
\end{eqnarray}
The canonical Hamiltonian density for this system has the next form 
\begin{eqnarray}
{\mathcal{H}}_{c}&=& \dot{e}{^{\mu}}_{I}  \Pi{_{I}}^{\mu}+ \dot{B}{^{I}}_{0a}\Pi{_{I}}^{0a}+  \dot{B}{^{I}}_{ab}\Pi{_{I}}^{ab}- {\mathcal{L}} \nonumber \\
&=& -  \frac{1}{2} \eta^{abc} B_{I0a} (\partial_b e{^{I}}_{c}-\partial_c e{^{I}}_b ) +\partial_ae{^{I}}_0\Pi{_{I}}^{a}. 
\label{eq11}
\end{eqnarray}
Integrating by parts  and neglecting boundary terms at infinity,  the canonical Hamiltonian  becomes 
\begin{equation}
H_c= \int dx^3 \left[ -  \frac{1}{2} \eta^{abc} B_{I0a} (\partial_b e{^{I}}_{c}-\partial_c e{^{I}}_b ) - \partial_a\Pi{_{I}}^{a}e{^{I}}_0 \right].
\label{eq12}
\end{equation}
Following with the method, adding to $H_{c}$ the 40 primary constraints (\ref{eq10}) we identify the primary Hamiltonian 
\begin{equation}
H_P= H_{c} +\int dx^3 \left [  \lambda^I{_0} \phi{_{I}}^{0}+\lambda^I{_a} \phi{_{I}}^{a}+\lambda^I{_{0a}}\phi{_{I}}^{0a}+\lambda^I{_{ab}} \phi{_{I}}^{ab} \right], 
\label{eq13}
\end{equation}
where  $\lambda^I{_0}, \lambda^I{_a}, \lambda^I{_{0a}}, \lambda^I{_{ab}}$ are Lagrange multipliers enforcing the constraints. For this theory, the non-vanishing  fundamental Poisson brackets are given by
\begin{eqnarray}
\{e^I{_{\alpha}}(x),  \Pi{_{J}}^{\mu}(y)  \} & =& \delta^\mu_\alpha \delta^I_J \delta^3(x-y), \nonumber \\
\{ B{^{I}}_{\mu \nu}(x), \Pi{_{J}}^{\alpha \beta}(y) \} &=& \frac{1}{2}\delta^I_J \left( \delta^\alpha_\mu \delta^\beta_\nu - \delta^\beta_\mu \delta^\alpha_\nu \right) \delta^3(x-y). 
\label{eq14}
\end{eqnarray}
The $40\times40$ matrix whose entries are the Posson brackets  among the constraints  (\ref{eq10}) given by 
\begin{eqnarray}
\{ \phi{_{I}}^{0}(x),\phi{_{J}}^{0}(y) \}&=&0,   \qquad   \{ \phi{_{I}}^{0}(x),\phi{_{J}}^{a}(y) \} = 0  \nonumber \\
\{ \phi{_{I}}^{0}(x),\phi{_{I}}^{0a}(y) \} &=& 0, \qquad  \{ \phi{_{I}}^{0}(x),\phi{_{I}}^{ab}(y) \} = 0, \nonumber \\
 \{ \phi{_{I}}^{a}(x), \phi{_{J}}^{b}(y)\} &=& 0,  \qquad  \{ \phi{_{I}}^{a}(x), \phi{_{J}}^{0b}(y)\} = 0, \nonumber \\
\qquad \qquad \qquad \{\phi{_{I}}^{0a} (x),\phi{_{J}}^{0b}(y) \} &=& 0, \qquad  \{ \phi{_{I}}^{a}(x), \phi{_{J}}^{cd}(y)\} = - \eta^{acd} \eta_{IJ} \delta^3(x-y) \nonumber \\
 \{\phi{_{I}}^{0a} (x),\phi{_{J}}^{cd}(y) \} &=& 0, \qquad  \{\phi{_{I}}^{ab} (x),\phi{_{J}}^{cd}(y) \} = 0 
 \label{eq15}
\end{eqnarray}
has rank 24 and 16 linearly independent null-vectors. Thus, the null vectors and  consistency conditions  yields to the next 16 secondary constraints \cite{19}
\begin{eqnarray}
\dot{\phi}{_{I}}^{0}&=& \{\phi{_{I}}^{0}, {\mathcal{H}}_{P} \} \approx 0 \quad \Rightarrow \quad \psi_I:= \partial_a\Pi{_{I}}^{a} \approx 0,  \nonumber \\ 
\dot{\phi}{_{I}}^{0a} &=& \{\phi{_{I}}^{0a},  {\mathcal{H}}_{P}  \} \approx0 \quad \Rightarrow \quad  \psi_I{^a}:= \frac{1}{2} \eta^{abc} (\partial_b e_{Ic}-\partial_c e_{Ib} ) \approx 0, 
\label{eq16}
\end{eqnarray}
and  the next values for the  Lagrange multipliers 
\begin{eqnarray}
\dot{\phi}{_{I}}^{a}&=&\{\phi{_{I}}^{a},  {\mathcal{H}}_{T}  \} \approx 0 \quad \Rightarrow \quad  \lambda^I{_{ab}}= \frac{1}{2}(\partial_a B{^{I}}_{0b} -\partial_b B{^{I}}_{0a}), \nonumber \\
\dot{\phi}{_{I}}^{ab}&=& \{ \phi{_{I}}^{ab}, {\mathcal{H}}_{T} \} \approx 0 \quad \Rightarrow \quad   \lambda^I{_a}=0, 
\label{eq17}
\end{eqnarray}
for the theory under study there are no, third  constraints.  At this point,  we need to separate all the  primary and secondary constraints in first and second class constraints. For this step,  we need   calculate the $56\times56$  matrix whose entries will be the Poisson brackets between primary and secondary constraints (\ref{eq9}) , (\ref{eq14}), this is
\begin{eqnarray}
\{ \phi{_{I}}^{0}(x),\phi{_{J}}^{0}(y) \}&=&0, \qquad \{ \phi{_{I}}^{0}(x),\phi{_{J}}^{a}(y) \} = 0, \nonumber \\
\{ \phi{_{I}}^{0}(x),\phi{_{I}}^{0a}(y) \} &=& 0, \qquad \{ \phi{_{I}}^{0}(x),\phi{_{I}}^{ab}(y) \} = 0, \nonumber \\
\{ \phi{_{I}}^{0}(x),\psi_J(y) \} &=& 0, \qquad \{ \phi{_{I}}^{0}(x),\psi_J{^a}(y) \} = 0, \nonumber \\
\{ \phi{_{I}}^{a}(x), \phi{_{J}}^{b}(y)\} &=& 0, \qquad \{ \phi{_{I}}^{a}(x), \phi{_{J}}^{0b}(y)\} = 0, \nonumber \\
\{ \phi{_{I}}^{a}(x), \psi_J(y)\} &=& 0, \qquad \{ \phi{_{I}}^{a}(x), \phi{_{J}}^{cd}(y)\} = - \eta^{acd} \eta_{IJ} \delta^3(x-y), \nonumber \\
\{ \phi{_{I}}^{a}(x), \psi_J(y)\} &=& 0, \qquad \{ \phi{_{I}}^{a}(x), \psi_J{^b}(y)\} =  -\eta^{abc} \eta_{IJ} \partial_c\delta^3(x-y), \nonumber \\
\{\phi{_{I}}^{0a} (x),\phi{_{J}}^{0b}(y) \} &=& 0, \qquad \{\phi{_{I}}^{0a} (x),\phi{_{J}}^{cd}(y) \} = 0, \nonumber \\
\{\phi{_{I}}^{0a} (x),\psi_J(y) \} &=& 0, \qquad \{\phi{_{I}}^{0a} (x),\psi_J{^b}(y) \} = 0, \nonumber \\
\{\phi{_{I}}^{ab} (x),\phi{_{J}}^{cd}(y) \} &=& 0, \qquad \{\phi{_{I}}^{ab} (x),\psi_J(y) \} = 0, \nonumber \\
\{\phi{_{I}}^{ab} (x),\psi_J{^c}(y) \} &=& 0, \qquad \{\psi_I (x),\psi_J(y) \} = 0, \nonumber \\
\{\psi_I (x),\psi_J{^a}(y) \} &=& 0, \qquad \{\psi_I{^a}(x),\psi_J{^b}(y) \} = 0,
\label{eq18}
\end{eqnarray}
this matrix has rank 24 and 32 null-vectors. Thus,  we expect  24 second class constraints and 32 first class constraints. From  the  null-vectors we identify  the next 32 first class constraints 
\begin{eqnarray}
\gamma_I{^0}&:=&  \Pi{_{I}}^{0} \approx 0 \nonumber \\
\gamma{_{I}}^{0a}&:=& \Pi{_{I}}^{0a} \approx 0,  \nonumber \\
\gamma_I&:=& \partial_a\Pi{_{I}}^{a} \approx 0,  \nonumber \\ 
\gamma_I{^a}&:=& \frac{1}{2} \eta^{abc} (\partial_b e_{Ic}-\partial_c e_{Ib} ) - \partial_b \Pi{_{I}}^{ab}\approx 0, 
\label{eq19}
\end{eqnarray}
and the rank yields to the  next 24 second class constraints 
\begin{eqnarray}
\chi{_{I}}^{a}&:=& \Pi{_{I}}^{a} - \eta^{abc}B_{I bc} \approx 0, \nonumber \\
\chi{_{I}}^{ab}&:=& \Pi{_{I}}^{ab} \approx 0.
\label{eq20}
\end{eqnarray}
It is important to remark that the constraint $\gamma_I{^a}$ given in (\ref{eq19}) is fixed by means of the null vectors (see equation (\ref{eq16})) and  become to be a first class constraint. In this way, the method itself allows us to find from   the rank and the null vectors  of the matrix (\ref{eq18})  all the right first  and second class constraints for the theory \cite{19}. This is the advantage  that we find in  Dirac's method when we apply it to the original configuration space, in this case given by $B^I{_{\alpha \beta}}$ and $e^I{_{\alpha}}$. In general we can apply the analysis presented in this work to every theory. However,   the calculation of the rank and the null vectors of the matrixes  (\ref{eq15}) and (\ref{eq18}) usually  is not straightforward  to perform \cite{19}. \\
Furthermore, the  32 first class constraints given in (\ref{eq19})  are not independent  because there are 4 reducibility conditions given by  $\partial_a \gamma_I{^a}= \partial_a \partial_b \chi_I^{ab}=0$, this reducibility condition is the equivalent  one that we find in the literature in the  4-dimentional BF theories \cite{20}  or in topological invariants  context  \cite{ 21a}.  In this manner,  the counting of degrees of freedom is a  follows.  There are   80 canonical variables $(e{^{I}}_{\mu}, B{^{I}}_{\alpha \beta}, \Pi{_{I}}^{\alpha}, \Pi{_{I}}^{\alpha \beta})$,   $ \left[ 32-4 \right]=28 $ independent first class constraints $(\gamma_I{^0}, \gamma{_{I}}^{0a},\gamma_I, \gamma_I{^a})$ and 24 independent second class constraints $(\chi{_{I}}^{a}, \chi{_{I}}^{ab})$, thus,  we can  conclude that theory is devoid of physical degrees of freedom. In others words, the theory defined by the action (\ref{eq5})  is only sensitive to external degrees of freedom for example,  if we add to (\ref{eq5}) matter degrees of freedom the theory will not be topological anymore, just  as was claimed in \cite{17}. In addition,  the action  (\ref{eq5})  does not depend explicit of the spacetime metric, so,  in this other sense the action becomes to be  topological as well \cite{20}. \\
With all these results at hand,  we can  use the values for the Lagrange multipliers (\ref{eq15}),  the first class constraints (\ref{eq19}), the second class constraints (\ref{eq20}) and identify  the extended action for the theory expressed by 
\begin{eqnarray}
&S_{E}&\left[ e{^I}_{\mu}, \Pi{_{I}}^\mu, B^I{_{\mu \nu}},  \Pi{_{I}}^{\mu \nu}, u{_{0}}^{I}, u^I, u{_{0a}}^{I},u{_{a}}^{I}, v{_a}^I, v^I{_{ab}} \right] 
= \int \bigg\{ \dot{e}^I{_\mu} \Pi{_I}^\mu+ \dot{B}^I{_{0a}}\Pi{_{I}}^{0a} +  \dot{B}^I{_{ab}}\Pi{_{I}}^{ab}\nonumber \\ &-& H- u{_{0}}^{I}\gamma{_{I}}^{0}-u^I \gamma_I- u^I{_a}\gamma{_{I}}^{a}-u{_{0a}}^{I}\gamma{_{I}}^{0a} - v{_a}^I\chi{_{I}}^{a}-v^I{_{ab}} \chi{_{I}}^{ab} \bigg\}dx^4,
\label{eq21}
\end{eqnarray}
where $H$ is only  combination of first class constraints 
\begin{equation}
H= - B^I{_{0a}}\left[ \frac{1}{2} \eta^{abc}(\partial_b e_{Ic}-\partial_c e_{Ib} ) -\partial_b\Pi{_{I}}^{ab} \right ]- 	\partial_a  \Pi{_{I}}^{a} e^I{_{0}}, 
\label{eq22}
\end{equation}
and $u{_{0}}^{I}, u^I, u{_{0a}}^{I},u{_{a}}^{I}, v{_a}^I, v^I{_{ab}}$ are the Lagrange multipliers enforcing the first and second class constraints. \\
From the extended action we can identify the extended Hamiltonian which is given by 
\begin{equation}
H_E= H - u{_{0}}^{I}\gamma{_{I}}^{0}-u^I \gamma_I- u^I{_a}\gamma{_{I}}^{a}-u{_{0a}}^{I}\gamma{_{I}}^{0a}. 
\label{eq23}
\end{equation}
As we know, the equations of motion obtained by means of the extended Hamiltonian in general are quite different with the Euler-Lagrande equations, but the difference is unphysical \cite{19}. \\
In oder to  complete our analysis, we can find  the equations of motion obtained from the extended action which yields to  
\begin{eqnarray}
\delta e^I{_{0}}:  \dot{\Pi}{_{I}}^{0}&=&- \partial_a\Pi{_{I}}^{a},  \nonumber \\
\delta \Pi{_{I}}^{0}:  \dot{e}^I{_{0}} &=& u^I{_{0}},  \nonumber \\
\delta e^I{_{a}}: \dot{\Pi}{_{I}}^{a}&=& - \frac{1}{2}  \eta^{abc} \left( \partial_b B_{I0c}- \partial_c B_{I0b} \right) -\frac{1}{2} \eta^{abc} \left( \partial_b u_{Ia}-\partial_c u_{Ib} \right) , \nonumber \\
\delta \Pi{_{I}}^{a}:   \dot{e}^I{_{a}}&=& v{_a}^I - \partial_a e^I{_{0}} - \partial_a u^I{_a}, \nonumber \\
\delta B^I{_{0a}}:  \dot{\Pi}{_{I}}^{0a}&=& \frac{1}{2} \eta^{abc}(\partial_b e_{Ic}-\partial_c e_{Ib} ) -\partial_b\Pi{_{I}}^{ab},  \nonumber \\
\delta \Pi{_{I}}^{0a}:  \dot{B}^I{_{0a}} &=&  u^I{_{0a}},  \nonumber \\
\delta B^I{_{ab}} : \dot{\Pi}{_{I}}^{ab}&=& \eta^{abc} v_{Ic}, \nonumber \\
\delta \Pi{_{I}}^{ab}:  \dot{B}^I{_{ab}}&=&  v^I{_{ab}} +  \frac{1}{2}(\partial_b B^I{_{0b}}-\partial_c B^I{_{0a}} ) - \frac{1}{2}(\partial_b u^I{_{b}}-\partial_c u^I{_{a}} )  \nonumber \\
\delta u{_{0}}^I:  \gamma{_{I}}^{0}&=&0, \nonumber \\
\delta u{_{a}}^I:  \gamma{_{I}}^{a}&=&0, \nonumber \\
\delta u^I:  \gamma{_{I}}&=&0, \nonumber \\
\delta u{_{0a}}: \gamma{_{I}}^{0a}&=&0, \nonumber \\
\delta v{_{a}}^I:  \chi{_{I}}^{a}&=&0, \nonumber \\
\delta v^I{_{ab}}: \chi{_{I}}^{ab}  &=&0.
\end{eqnarray}
On the other hand, we will calculate the constraint algebra which takes the form 
\begin{eqnarray}
\{ \gamma{_{I}}^{0}(x),\gamma{_{J}}^{0}(y) \}&=&0, \qquad \{ \gamma{_{I}}^{0}(x),\chi{_{J}}^{a}(y) \} = 0, \nonumber \\
\{ \gamma{_{I}}^{0}(x),\gamma{_{I}}^{0a}(y) \} &=& 0, \qquad \{ \gamma{_{I}}^{0}(x),\chi{_{I}}^{ab}(y) \} = 0, \nonumber \\
\{ \gamma{_{I}}^{0}(x),\gamma_J(y) \} &=& 0, \qquad \{ \gamma{_{I}}^{0}(x),\gamma_J{^a}(y) \} = 0, \nonumber \\
\{ \chi{_{I}}^{a}(x), \gamma{_{J}}^{b}(y)\} &=& 0, \qquad \{\chi{_{I}}^{a}(x), \gamma{_{J}}^{0b}(y)\} = 0, \nonumber \\
\{ \chi{_{I}}^{a}(x), \gamma_J(y)\} &=& 0, \qquad \{ \chi{_{I}}^{a}(x), \chi{_{J}}^{cd}(y)\} = - \eta^{acd} \eta_{IJ} \delta^3(x-y), \nonumber \\
\{ \chi{_{I}}^{a}(x), \gamma_J(y)\} &=& 0, \qquad \{ \chi{_{I}}^{a}(x),\gamma_J{^b}(y)\} =  0, \nonumber \\
\{\gamma{_{I}}^{0a} (x),\gamma{_{J}}^{0b}(y) \} &=& 0, \qquad \{\gamma{_{I}}^{0a} (x),\gamma{_{J}}^{cd}(y) \} = 0, \nonumber \\
\{\gamma{_{I}}^{0a} (x),\gamma_J(y) \} &=& 0, \qquad \{\gamma{_{I}}^{0a} (x),\gamma_J{^b}(y) \} = 0, \nonumber \\
\{\chi{_{I}}^{ab} (x),\chi{_{J}}^{cd}(y) \} &=& 0, \qquad \{\chi{_{I}}^{ab} (x),\gamma_J(y) \} = 0, \nonumber \\
\{\chi{_{I}}^{ab} (x),\gamma_J{^c}(y) \} &=& 0, \qquad \{\gamma_I (x),\gamma_J(y) \} = 0, \nonumber \\
\{\gamma_I (x),\gamma_J{^a}(y) \} &=& 0, \qquad \{\gamma_I{^a}(x),\gamma_J{^b}(y) \} = 0,
\end{eqnarray}
where we can see that  the constraint algebra  is closed. \\
We will finish  this section   identify the Dirac bracket for the theory. From  the constraint algebra,   we can observe  that  the  matrix whose elements are only the Poisson brackets between the second class constraints  is given by 
\begin{equation}
C_{\alpha\beta} = \left(
\begin{array}{rr}
0 \qquad \qquad& - \eta^{acd} \eta_{IJ} \delta^3(x-y)  \\
 \eta^{acd} \eta_{IJ} \delta^3(x-y) & 0 \qquad \qquad \\
 \label{eqa}
\end{array}
\right).
\end{equation}
In this manner,  we have  that the Dirac bracket between   two functionals $A$, $B$  is expressed   by 
\begin{equation}
\{A(x),B(y) \}_D= \{A(x),B(y)\}_P + \int du dv \{A(x), \zeta^\alpha(u) \} C^{-1}_{\alpha \beta}(u,v) \{\zeta^\beta(v), B(y) \}, 
\label{eq27}
\end{equation}
where $ \{A(x),B(y)\}_P$ is the usual Poisson bracket between the functionals $A,B$,   $\zeta^\alpha(u)=(\chi{_{I}}^{a}, \chi{_{I}}^{ab} ) $ with   $C^{-1}_{\alpha \beta}(u,v)$  as  the inverse of (\ref{eqa}) which has a  trivial form. As we know, the Dirac bracket (\ref{eq27}) will be useful  to make progress in the quantization of the theory.\\
\newline
\newline
\noindent \textbf{II.I  Gauge generator}\\[1ex]
Following with the method, in this part  we will find the gauge transformations for the theory described by (\ref{eq5}). For our purposes,  we  apply  the Castellani's algorithm  \cite{21} to construct the gauge generator  using the first class constraints (\ref{eq19}), this is
\begin{eqnarray}
G= \int_\Sigma \left[ \partial_0 \varepsilon^I{_{0}} \Pi{_{I}}^{0} + \partial_0 \varepsilon^I{_{0a}}\Pi{_{I}}^{0a}+\varepsilon^I \partial_a \Pi{_{I}}^{a} + \varepsilon^I{_{a}}\left(   \frac{1}{2} \eta^{abc} (\partial_b e_{Ic}-\partial_c e_{Ib} ) - \partial_b \Pi{_{I}}^{ab}\right)  \right], 
\label{eq24}
\end{eqnarray}
thus, we find the following gauge transformations on the phase  space, 
\begin{eqnarray}
\delta_0 e^I{_{0}} &=& \partial_0  \varepsilon^I{_{0}},  \nonumber \\
\delta_0 e^I{_{a}} &=& -  \partial_a  \varepsilon^I,  \nonumber \\
\delta_0 B^I{_{0a}} &=&   \partial_0  \varepsilon^I{_{0a}},  \nonumber \\
\delta_0 B^I{_{ab}} &=& -\frac{1}{2}  (\partial_a \varepsilon^I{_b} - \partial_b \varepsilon^I {_a}),  \nonumber \\
\delta_0 \Pi{_{I}}^{0}   &=&   0,  \nonumber \\
\delta_0 \Pi{_{I}}^{a}   &=&  -\frac{1}{2} \eta^{abc}  (\partial_b \varepsilon{_{Ic}} - \partial_c \varepsilon{_{Ib}})  ,  \nonumber \\
\delta_0 \Pi{_{I}}^{0a}  &=&   0,  \nonumber \\
\delta_0 \Pi{_{I}}^{ab}  &=&   0.  
\label{eq25}
\end{eqnarray}
In particular,  we can choose   the parameters to be $ \varepsilon^I{_{0}}=- \varepsilon^I=-\Lambda^I$,  $\varepsilon^I{_{a}}=-2\varepsilon^I{_{0a}}=\Lambda^I{_{a}}$ and considering the equations (\ref{eq25}) we find 
\begin{eqnarray}
e^I{_{\mu}}& \rightarrow& e^I{_{\mu}}- \partial_\mu\Lambda^I, \nonumber \\
B^I{_{\mu \nu}}& \rightarrow& B^I{_{\mu \nu}}- \frac{1}{2} \left(\partial_\mu \Lambda^I{_{\nu}} -\partial_\nu \Lambda^I{_{\mu}}  \right), \nonumber \\
\label{eq26}
\end{eqnarray}
where we can see that $e^I{_{\mu}}$ becomes to be a collection of 4 four gauge invariant vector fields. We can prove by means of easy calculations that the action (\ref{eq5}),  the equations of motion (\ref{eq3}) and  (\ref{eq4}) are invariant  under these   gauge  transformations. The nature of the gauge transformations and the form of the  theory described in (\ref{eq5}) which corresponds to  $BF$ type, allows us to formulate the next question;  What about  diffeomorphisms transformations?. Apparently  diffeomorphisms symmetry is not present in the theory, but   that is not true at all.  We can find the answer such as is developed in  2+1 gravity and Chern-Simons theory  \cite{21, 23} introducing  a new set of gauge parameters 
\begin{eqnarray}
\Lambda^I&=&- \xi^\rho e^I{_\rho}, \nonumber \\
 \Lambda^I{_{\mu}} &=& -2\xi ^\rho  B^I{_{\rho \mu}}, 
\end{eqnarray} 
obtaining 
\begin{eqnarray}
e^I{_{\mu}}& \rightarrow& e^I{_{\mu}}+  {\mathcal{L}}_\xi  e^I{_{\mu}} + \xi^\rho \left[ \partial_\mu e^I{_{\rho}}-\partial_\rho e^I{_{\mu}}  \right], \nonumber \\
B^I{_{\mu \nu}}& \rightarrow& B^I{_{\mu \nu}} +  {\mathcal{L}}_\xi B^I{_{\mu \nu}} + \xi^\rho \left[\partial_\mu  B^I{_{\rho \nu}}-\partial_\nu  B^I{_{\rho \mu}}- \partial_\rho B^I{_{\mu \nu}}  \right].
\label{eq28}
\end{eqnarray}
Therefore,  diffeomorphisms  corresponds to  an  internal symmetries  of  the theory just as complete general relativity theory. \\
As conclusion for this section,  we can see that it is possible to obtain all the physical information reported in \cite{17} without resort to $ADM$ variables. Of course, we can obtain the results obtained in \cite{17} considering the second class constraints given in (\ref{eq20}) as strong equations. However,  the spirit of this paper is make progress for futures works where we will investigate the advantage  at quantum  level between the  $ADM$  formulation and the formulation presented in this work.\\
\newline
\newline
\noindent \textbf{III  Covariant canonical formalism}\\[1ex]
In order to  extend  our analysis, in this section we will perform the covariant canonical formalism for the theory described by the action  (\ref{eq5}). In particular   with this method  we will establish the necessary elements  for  study the quantization aspects of  the theory in future works, where   we will  use the symplectic method  or the Hamiltonian method developed above. As important results reported in this section,  we will find by other way  the symmetries  found using  the Hamiltonian method. \\ 
We start calculating the variation of the action,  obtaining 
\begin{equation}
\delta S[B, e]=  \int_M  dx^4 \left[ \frac{1}{2} \epsilon^{\alpha \beta \mu \nu}(\partial_\mu e_{\nu I}- \partial_{\nu}e_{\mu I}) \delta B{^{I}}_{\alpha \beta }  - \epsilon^{\alpha \beta \mu \nu} \partial_\mu B{^{I}}_{\alpha \beta }  \delta e^I{_\nu}+ \partial_\mu(\epsilon^{\alpha \beta \mu \nu}B_{I \alpha \beta } \delta e^I{_{\nu}}  ) \right], 
\label{eq29}
\end{equation}
where we can identify the  equations of motion  (\ref{eq3}), (\ref{eq4}) and we identify from the pure divergence term  the symplectic potential for the theory  \cite{22}
\begin{equation}
\Psi^\mu = \epsilon^{ \mu \nu \alpha \beta}B_{I \alpha \beta } \delta e^I{_{\nu}}, 
\label{eq30}
\end{equation}
which does not  contribute locally to the dynamics, but generates the symplectic form  on the phase space.\\
From the equations of motion (\ref{eq3}) and  (\ref{eq4}) we define the fundamental concept in the studio of the covariant canonical formalism of the theory: the covariant phase space for the theory described by (\ref{eq5})  is the space space of solutions of Eqs (\ref{eq3}), (\ref{eq4}), and we will call it $Z$.\\
As we known, we can obtain  the integral kernel of the geometric structure for the theory  by  means of the variation (exterior derivative on $Z$ see \cite{22}) of the symplectic potential (\ref{eq30}),  this is 
\begin{equation}
\omega =  \int_\Sigma J^\mu d \Sigma_\mu=  \int_\Sigma  \delta \Psi^\mu d \Sigma_\mu= \int_\Sigma \epsilon^{ \mu \nu \alpha \beta}\delta B_{I \alpha \beta }\wedge  \delta e^I{_{\nu}} d \Sigma_\mu .
\label{eq31}
\end{equation}
where $\Sigma$ is a Cauchy hypersurface.\\  
In addition, we will  prove that our symplectic form  is closed and gauge invariant. Moreover,  the integral kernel of the geometric form $J^\mu$ is conserved $(\partial_\mu J^\mu=0)$, which guarantees that $\omega$  is independent of $\Sigma$. \\ To prove that  $J^\mu$ defined in (\ref{eq31}) is conserved   we need calculate the linearized equations of motion. For this, we replace in  (\ref{eq3}),   (\ref{eq4})   $ e^I{_{\nu}} \rightarrow e^I{_{\nu}}+ \delta e^I{_{\nu}} $ and $ B_{I \alpha \beta }  \rightarrow B_{I \alpha \beta }+\delta B_{I \alpha \beta }$,  keeping to first order in $\delta$ we find  the linearized equations given by 
\begin{eqnarray}
\epsilon^{\alpha \beta \mu \nu }\partial_{[\mu} \delta e_{\nu]I} &=&0, \nonumber \\
 \epsilon^{\alpha \beta \mu \nu } \partial_\mu \delta  B_{I \alpha \beta}&=&0.
 \label{eq32}
\end{eqnarray}
In this manner, using the linearized equations we have 
\begin{equation}
\partial_\mu J^\mu= \partial_\mu \delta \Psi^\mu = \epsilon^{ \mu \nu \alpha \beta} \partial_\mu \delta B_{I \alpha \beta } \wedge \delta e^I{_{\nu}}+\epsilon^{ \mu \nu \alpha \beta}\delta B_{I \alpha \beta } \wedge \partial_{[\mu} \delta  e^I{_{\nu]}}=0, 
\label{eq33}
\end{equation}
showing that  $\omega $ is independent of $\Sigma$. \\
On the other hand, we need to remember that  the closeness of $\omega$ in this covariant  canonical formalism is equivalent one  to the Jacobi identity that Poisson brackets satisfy, in the usual  Hamiltonian scheme.  To prove the closeness of $\omega$, we can  observe that $\delta^2e^I{_{\nu}}=0 $, $\delta^2B_{I \alpha \beta }=0$ because $e^I{_{\nu}}$ and  $B_{I \alpha \beta }$ are independent 0-forms on the covariant phase space $Z$ and  $\delta$ is nilpotent, so  using this fact in $\omega$ we find 
\begin{equation}
\delta \omega =\int _\Sigma \delta^2 \Psi^\mu d \Sigma_\mu  = \int _\Sigma \left[  \epsilon^{ \mu \nu \alpha \beta}\delta^2 B_{I \alpha \beta }\wedge  \delta e^I{_{\nu}}-\epsilon^{ \mu \nu \alpha \beta}\delta B_{I \alpha \beta }\wedge  \delta^2 e^I{_{\nu}} \right] d \Sigma_\mu =0, 
\end{equation}
this prove that $\omega $ is closed. \\
What about the gauge transformations found   above?.  For this aim, we consider that upon picking $\Sigma$ to be the standard initial value surface $t=0$, (\ref{eq31}) takes the standard form 
 \begin{equation}
\omega = \int_\Sigma \delta  \Pi{_{I}}^{a}\wedge  \delta e^I{_{a}}, 
\label{eq33a}
\end{equation}
where $ \Pi{_{I}}^{a}\equiv \eta^{ abc} B_{I bc }$. \\
For two 0-forms $f, g$ defined on $Z$,  the Hamiltonian vector field defined by the symplectic structure (\ref{eq33a}) is given by \cite{24}
\begin{equation}
X_f= \int_{\Sigma} \frac{\delta f}{\delta \Pi{_{I}}^{a}} \frac{\delta }{\delta e^I{_{a}}}- \frac{\delta f}{\delta e^I{_{a}}}  \frac{\delta }{\delta \Pi{_{I}}^{a}}, 
\label{eq33b}
\end{equation}
and the Poisson bracket $\{f,g \}:= -X_f(g)$ is given by 
\begin{equation}
\{f,g \}= \int_{\Sigma} \frac{\delta f}{\delta e^I{_{a}}}  \frac{\delta g}{\delta \Pi{_{I}}^{a}}-  \frac{\delta f}{\delta \Pi{_{I}}^{a}} \frac{\delta g }{\delta e^I{_{a}}}.
\label{eq33c}
\end{equation}
On the other hand, we rewrite the first class constraints found in (\ref{eq19}) with the test fields $D^I, D^I{_{a}}, C^I $ and $C^I{_{a}}$ on $\Sigma$ in the next form
\begin{eqnarray}
\gamma_I{^0}[D^I]&:=&\int_{\Sigma} D^I  \left( \Pi{_{I}}^{0} \right), \nonumber \\
\gamma{_{I}}^{0a}[D^I{_{a}}]&:=&\int_{\Sigma}D^I   \left(  A^I{_{a}} \Pi{_{I}}^{0a} \right),  \nonumber \\
\gamma_I[C^I ]&:=& \int_{\Sigma} C^I  \left(  \partial_a\Pi{_{I}}^{a}\right),  \nonumber \\ 
\gamma_I{^a}[C^I{_{a}}]&:=& \int_{\Sigma} C^I{_{a}} \left( \frac{1}{2} \eta^{abc} (\partial_b e_{Ic}-\partial_c e_{Ib} ) - \partial_b \Pi{_{I}}^{ab} \right). 
\label{eq33d}
\end{eqnarray}
By inspection, the functional derivatives  different to zero are given by 
\begin{eqnarray}
\frac{\delta \gamma_I[C^I ]}{\delta \Pi{_{I}}^{a} }&=& - \partial_a C^I, \quad \quad \quad \frac{\delta \gamma_I[C^I ]}{ \delta e^I{_{a}}}=0, \nonumber \\
\frac{\delta \gamma_I{^a}[C^I{_{a}}]}{\delta \Pi{_{I}}^{a} }&=& 0, \quad \quad \quad \frac{\delta \gamma_I{^a}[C^I{_{a}}]}{ \delta e^I{_{a}}}= \frac{1}{2}\eta^{abc}\left(\partial_b C_{Ic}-\partial_c C_{Ib}  \right).
\label{eq33f}
\end{eqnarray}
Thus, the motion on $Z$ generated  by  $\gamma_I[C^I ]$ is given by 
\begin{eqnarray}
e^I{_{a}} &\mapsto& e^I{_{a}}  - \epsilon \partial_a C^I + O(\epsilon^2) \nonumber \\
\Pi{_{I}}^{a}  &\mapsto& \Pi{_{I}}^{a}, 
\label{eq33g}
\end{eqnarray}
and the motion on $Z$ generated by $ \gamma_I{^a}[C^I{_{a}}]$ is given by 
\begin{eqnarray}
e^I{_{a}} &\mapsto& e^I{_{a}}  \nonumber \\
\Pi{_{I}}^{a}  &\mapsto& \Pi{_{I}}^{a}-\epsilon \frac{1}{2}\eta^{abc}\left(\partial_b C_{Ic}-\partial_c C_{Ib}  \right) + O(\epsilon^2). 
\label{eq33h}
\end{eqnarray}
where $\epsilon$ is an infinitesimal parameter \cite{24}. We can see that the gauge transformation (\ref{eq33g}) and (\ref{eq33h}) corresponds to those  found using   Dirac's method (see eq. (\ref{eq26}) ).\\
Now, we will  show that $\omega$ has not components tangent to the gauge directions, which are specified  by   equation (\ref{eq26}) or (\ref{eq33g}) and (\ref{eq33h}).
\begin{eqnarray}
\delta e' {^I}{_{\mu}}& =& \delta e^I{_{\mu}}- \partial_\mu\Lambda^I, \nonumber \\
\delta B' {^I}{_{\mu \nu}}& =& \delta B^I{_{\mu \nu}}- \frac{1}{2} \left(\partial_\mu \Lambda^I{_{\nu}} -\partial_\nu \Lambda^I{_{\mu}}  \right),
\label{eq33} 
\end{eqnarray}
where in this context $\Lambda^I$ , $\Lambda^I{_{\mu}}$ corresponds to be  1-forms on $Z$.  Using this fact,  we find that $\omega$ will undergo the transformation as 
\begin{equation}
\omega '  = \int_\Sigma \epsilon^{ \mu \nu \alpha \beta}\delta B' _{I \alpha \beta }\wedge  \delta e' {^I}{_{\nu}} d \Sigma_\mu= \omega - \int_\Sigma   \partial_\nu \left[\frac{1}{2} \epsilon^{ \mu \nu \alpha \beta} \left( \partial_\alpha \Lambda{_{I\beta}} -\partial_\beta \Lambda{_{I \alpha}} \right) \wedge \Lambda^I \right]d \Sigma_\mu, 
\end{equation}
where  the equations  (\ref{eq32}) has been used, thus, for fields with compact support  $\omega$ is a gauge invariant geometric form.\\
Therefore, as a conclusion of this section, we have constructed a closed and gauge invariant symplectic form on $Z$  which in turns represent a complete Hamiltonian description of the covariant phase space for the theory and will  allow us to analyze the quantum treatment in forthcoming works.\\
\newline
\newline
\newline
\noindent \textbf{V. Conclusions and prospects}\\[1ex]
 In this paper,  Dirac  and the symplectic  methods for the Einstein's action in the  $G \rightarrow0 $ limit   has been performed.  Within the Dirac's method we developed the analysis working with the complete  configuration space and without involve the typical   $ADM$ variables  as is reported  in \cite{17}. As important results obtained using the Hamiltonian method,    were the  identification of   the extended Hamiltonian, the extended action and   the separation  of the constraints in  first and second class. The correct identification  of the constraints allowed  us to find  the relevant symmetries  such as  the diffeomorphisms and could carry out the  counting of the physical degrees of freedom,  which the analysis allow one to conclude  that the system  is a topological  field theory. It is important to remark that the present analysis  can be useful to understand  the $G \rightarrow0 $ limit of general relativity, because we have present  a  background independent and  full diffeomorphism invariant free field theory. This fact becomes to be important because in the analysis  we have not broken  the important symmetries that characterize to Eintein's theory of gravity.  In addition,  we extended  our work  constructing  a closed and gauge invariant  symplectic structure which  contains all the  relevant Hamiltonian description of the covariant phase space. In particular using the geometric form,  we could find  the same symmetries  that we found  using the Hamiltonian method.  With the results presented in this paper, we have all the necessary elements to make progress in the quantization of the theory by means of the Dirac's method or  covariant canonical formalism which is absent in the literature and will be reported  in forthcoming works. \\
\newline
\newline
\newline
\noindent \textbf{Acknowledgements}\\[1ex]
This work was supported by CONACyT  M\'exico under grant 76193.  I want to thank to  Brandon Carter  and Eric Gourgouhon  for  the hospitality and  friendship that they  have offered me. 

\end{document}